\documentclass[a4paper]{jpconf}
\usepackage{amsmath,amsfonts,amssymb}
\usepackage{graphicx,hyperref}

\newcommand*{\cf}{cf.\ }
\newcommand*{\ie}{i.e.\ }
\newcommand*{\eg}{e.g.\ }
\newcommand*{\myRef}{ref.~}

\begin{document}
\hfill TTP21-060, P3H-21-101 \\
\title{FeynCalc goes multiloop}

\author{Vladyslav Shtabovenko}

\address{Institut für Theoretische Teilchenphysik (TTP), Karlsruhe Institute of Technology (KIT), Wolfgang-Gaede-Straße 1, 76131 Karlsruhe, Germany}

\ead{v.shtabovenko@kit.edu}

\begin{abstract}
We report on the new functionality of the open-source \textsc{Mathematica} package \textsc{FeynCalc} relevant for multiloop calculations. In particular, we focus on such tasks as topology identification by means of the Pak algorithm, search for equivalent master integrals and their graph representations as well as automatic derivations of Feynman parametric representations for a wide class of multiloop integrals. The functions described in this report are expected to be finalized with the official release of \textsc{FeynCalc} 10. The current development snapshot of the package including the documentation is publicly available on the project homepage. User feedback is highly encouraged.
\end{abstract}

\section{Introduction}
\vspace{0.2cm}
There is a widespread consensus among particle theorists that in order to match the expected experimental precision at the
High Luminosity Large Hadron Collider (HL-LHC) we need new ideas and algorithms for our analytical and numerical calculations. This is especially true for the field of multiloop computations (\cf \eg\cite{Tancredi:2021oiq} for an overview of recent developments), where analytic evaluations of hundreds, thousands and sometimes even millions of Feynman diagrams often require the practitioners to expand the borders of what is feasible using computer algebraic methods. Despite such remarkable advances that corroborate the healthy state of the field, one should not forget that breakthrough calculations with many loops or legs are far from being commonplace. In fact, the number of groups worldwide that possess the expertise, computational resources and \emph{software tools} to work at the forefront of the precision frontier is rather low. Especially the role of computer programs in this context should not be underestimated. The truth is that such codes are of utmost importance to the task of calculating multiloop amplitudes in a straightforward and efficient fashion. Unfortunately, for various reasons many of such powerful codes are not publicly available, which severely impedes the adoption of new and efficient computational techniques among particle phenomenologists. After all, the number of theorists that have time, motivation, diligence and programming skills to implement relevant algorithms single-handedly for the project at hand, is not that high. Most people, therefore, tend to rely on tools that are (i) freely available, (ii) well documented, (iii) easy to use and (iv) are actively supported by their developer teams.

One of such programs is the open-source \textsc{Mathematica} package \textsc{FeynCalc}~\cite{Mertig:1990an,Shtabovenko:2016sxi,Shtabovenko:2020gxv} that has been available to the community since more than three decades. \textsc{FeynCalc} is known as a tool that can be employed in highly nonstandard scenarios that require full control over each calculational step. For example, a rather unique feature of the package that has been added very recently, comprises the ability to automatize manifestly noncovariant calculations~\cite{Brambilla:2020fla}. Although one might argue that nonrelativistic quantum field theories is a very small niche, this functionality has been warmly embraced by the effective field theory (EFT) community and helped to obtain interesting new research results~\cite{Brambilla:2017kgw,Brambilla:2019fmu,Assi:2020srd,Assi:2020piz,Urban:2021evz,Biondini:2021ccr,Chung:2021efj,Frugiuele:2021bic,Biondini:2021ycj}

However, the common perception of \textsc{FeynCalc} is that it is useful only for tree-level and one-loop calculations, while everything at two loops and beyond should be handled using other tools. In this conference note we intend to challenge this viewpoint by reporting on selected capabilities of the upcoming \textsc{FeynCalc} 10~\cite{Shtabovenko:FCX}. More explicitly, we will discuss the usage of \textsc{FeynCalc} for the tasks of topology identification and manipulations of master integrals obtained after a successful Integration-By-Parts (IBP)~\cite{Chetyrkin:1981qh,Tkachov:1981wb} reduction. These two steps necessarily arise in almost any multiloop calculation and can be usually handled using \textsc{Mathematica} codes without hitting performance bottlenecks. Furthermore, they have been inspired by the author's multiloop-related research work~\cite{Dixon:2018qgp,Luo:2019nig,Gao:2020vyx,Gerlach:2021xtb} and thus at least partially bench-tested on real-life problems.

\section{Topology identification}
\vspace{0.2cm}
Topology identification is commonly understood as the procedure of assigning all loop integrals occurring in the given calculation to a set of integral families. An integral family or topology consists of a list of linearly independent propagators that form a basis. While the problem at hand may easily contain tens of thousands of integrals, the number of topologies they belong to is usually much smaller ranging from dozens to several hundreds.\footnote{Of course, the exact number highly depends on the considered process and the number of relevant scales.} Topology identification is also a necessary step for running IBP reduction using tools\footnote{\textsc{Reduze}~\cite{Studerus:2009ye,vonManteuffel:2012np} is somewhat special in this respect, as it has built-in topology identification routines.} such as \textsc{FIRE}~\cite{Smirnov:2019qkx}, \textsc{KIRA}~\cite{Maierhofer:2017gsa,Klappert:2020nbg} or \textsc{LiteRed}~\cite{Lee:2012cn,Lee:2013mka}.

The process of mapping integrals to topologies can be done on the level of graphs or analytic amplitudes. In the former case each of the given Feynman diagrams or amplitudes is converted to a graph, which transforms the initial problem to the task of finding subgraph isomorphisms. In the latter case one works directly with symbolic propagators and tries to identify sets that are equivalent upon applying suitable loop momentum shifts. Pure graph-based approach is realized in the \textsc{C++} programs \textsc{q2e/exp}~\cite{Harlander:1998cmq,Seidensticker:1999bb}, while \textsc{Mathematica} package \textsc{TopoID}~\cite{Hoff:2016pot} implements the analytic mapping procedure. Other codes such as \textsc{Reduze} (\textsc{C++})~\cite{Studerus:2009ye,vonManteuffel:2012np}, \textsc{feynson} (\textsc{C++})~\cite{url:Feynson}, \textsc{tapir} (\textsc{Python})~\cite{Gerlach:tapir} or \textsc{pySecDec}~\cite{Borowka:2017idc,Borowka:2018goh,Heinrich:2021dbf} combine multiple methods in creative ways.

In this report we would like to focus on the task of finding a set of unique integral topologies for the given amplitude. This amplitude can equally represent a single Feynman diagram or a sum thereof. The first step is always to look at all the occurring sets of loop integral denominators and try to detect which of them can be mapped into each other. At this stage we do not need to look at the numerators. This is because for a set of propagators forming a basis, each scalar product involving loop momenta can be always expressed as a linear combination of inverse propagators. 

Naively, one might try to search for equivalences between different topologies by applying all possible loop momentum shifts in the hope that a right combination of shifts will produce a match. While this procedure works in principle, the combinatorial growth of complexity effectively prohibits its application to most realistic multiloop problems. A much more efficient algorithm for this special problem was devised by Alexey Pak in~\cite{Pak:2011xt}. A very detailed and pedagogical summary of Pak's ideas can be found in the doctoral thesis of Jens Hoff~\cite{Hoff:2015kub}, which was enormously useful for the presented implementation.

The main idea behind the Pak algorithm is to compare topologies represented in form of the so-called graph or Symanzik polynomials $\mathcal{U}$ and $\mathcal{F}$ (\cf\cite{Bogner:2010kv} for an extensive review). These two quantities naturally arise when converting a loop integral into a Feynman parametric integral using the well-known formula (\cf \eg\cite{Smirnov:2012gma})

\begin{align}
	& \left (\frac{e^{\varepsilon \gamma_E}}{i \pi^{d/2}} \right )^L \int  \frac{\left ( \prod_{i=1}^L d^d k_i \right )}{P_1^{m_1} \ldots P_N^{m_N}} \nonumber \\ 	
	& =  \frac{(-1)^{N_m} \Gamma(N_m-\frac{L d}{2})}{\prod_{j=1}^N \Gamma(m_j)} \int_0^\infty \prod_{j=1}^N d x_j x_j^{m_j-1} \, \delta \left ( 1- \sum_{i=1}^N x_i  \right )  \frac{\mathcal{U}^{N_m-\frac{(L+1) d}{2}} }{\mathcal{F}^{N_m-\frac{L d}{2}}}
\end{align}
where $P_i$ denote quadratic or eikonal propagators and $N_m = \sum_{i=1}^N m_i$. Notice that this formula is valid only for positive propagator powers \ie $m_i\geq 0$. Nevertheless, it can be easily extended to negative powers and even tensor structures in the numerators (\cf \eg doctoral thesis of Stefan Jahn~\cite{Jahn:2020tpj} for a nice summary). From here one can start comparing topologies by looking at their \emph{characteristic polynomials} defined as $\mathcal{P} \equiv \mathcal{U} \times \mathcal{F}$. The usage of these polynomials removes ambiguities arising from momentum shifts that often make equivalent topologies seem very different. However, the given characteristic polynomial still cannot be regarded as a unique representation of the corresponding loop integral. This is because $\mathcal{P}$ is not invariant under arbitrary renamings of Feynman parameters (\eg $x_1 \to x_3$, $x_2 \to x_1$ etc.). Each such renaming yields a new polynomial that differs from the old one, even though both of them describe the same integral. Pak algorithm removes this ambiguity by introducing the notion of \emph{canonical oreding} for the given $\mathcal{P}$. This means that by comparing two canonical polynomials $\mathcal{P}_1$ and $\mathcal{P}_2$ one can unambiguously answer the question whether the corresponding topologies $T_1$ and $T_2$ are identical or not.

Let us now come to the implementation of this technology in \textsc{FeynCalc}. In order to avoid cluttering the text we will abstain from including code examples in the body of this report. Those examples can be found in the ancillary file accompanying this work.

First of all, \textsc{FeynCalc} 10 introduces two new symbols representing single loop integrals and loop integral topologies. These are \texttt{GLI} and \texttt{FCTopology}. For example, \texttt{GLI[topo, \{1,0,1,1,2\}]} is a placeholder for some \emph{generic loop integral} that belongs to the integral family \texttt{topo} containing 5 propagators. Our integral contains only 4 lines, with one of the propagators being raised to a quadratic power. \texttt{FCTopology}, on the contrary, features a richer structure that consists of 6 arguments as in \texttt{FCTopology[topo, \{propagators\}, \{loop momenta\}, \{external momenta\}, \{kinematics\}, \{\}]}. The first 5 arguments should be self-explanatory, while the last argument (an empty list) is reserved for additional information such as cuts or symmetries but is currently ignored by the existing functions. The majority of new multiloop-related routines expect their first two arguments to be a list of \texttt{GLI}s and a list of the corresponding \texttt{FCTopology} objects. In many cases it is also possible to input loop integrals in the traditional \texttt{(S)FAD}-notation that uses explicit $D$-dimensional scalar products (\texttt{SPD}s) and propagators (\texttt{FAD}s or \texttt{SFAD}s\footnote{The main difference between \texttt{FAD}s and \texttt{SFADs} is that the former can encode only quadratic propagators, while the latter allow for both quadratic and eikonal propagators \cf Sec.~4.3 in \cite{Brambilla:2020fla}.}). Furthermore, it is always possible to convert an integral in the \texttt{GLI}-notation to the \texttt{SFAD}-notation using \texttt{FCLoopFromGLI}.

We now want to proceed to the characteristic polynomials and the Pak algorithm. Traditionally,  many \textsc{FeynCalc} routines that perform complicated manipulations on symbolic expressions (\eg \texttt{TID} or \texttt{DiracSimplify}) are actually built on top of numerous auxiliary functions that are equally accessible to the user. This modular approach is one of the reasons why \textsc{FeynCalc} easily integrates into different workflows. Instead of being a black box that forces the user to follow specific paths and patterns, \textsc{FeynCalc} is designed to adapt itself to the user's needs. The multiloop extension of the package strictly follows this philosophy.

One of the most fundamental routines related to the handling of multiloop integrals is \texttt{FCFeynmanPrepare}. The main task of this function is to compute the $\mathcal{U}$ and $\mathcal{F}$ polynomials of the given integral using the algorithm from the famous \texttt{UF.m} program~\cite{Smirnov:2012gma} that is currently part of \textsc{FIESTA}~\cite{Smirnov:2021rhf}. In addition to that, \texttt{FCFeynmanPrepare} also returns other important building blocks such as the matrix $M$ with $\mathcal{U} = \det M$ as well as $J$ and $Q^\mu$ from $\mathcal{F} = \det M ( Q M^{-1} Q - J)$. These quantities are not really needed for the topology identification but can be very useful for other purposes \eg when constructing Feynman parametric representations.

The characteristic polynomial $\mathcal{P}$ can be obtained from \texttt{FCLoopToPakForm}, where we would like to stress that the polynomial returned by this function is already canonically ordered. Of course, it is also possible to determine the canonical ordering of a user-defined polynomial that does not necessarily has to be related to a particular loop integral. The corresponding function is called \texttt{FCLoopPakOrder}. It is worth noting that by analyzing $\mathcal{P}$ it is possible to detect scaleless loop integrals that vanish in dimensional regularization. This also concerns cases where  one cannot readily recognize the scalelessness by merely looking at the integral. Here we refer to Sec.~2.3 of \myRef\cite{Hoff:2015kub} for a description of the underlying algorithm. This functionality is implemented in \textsc{FeynCalc} in form of the functions \texttt{FCLoopPakScalelessQ} (for characteristic polynomials) and \texttt{FCLoopScalelessQ} (for loop integrals).

The actual task of performing the topology identification is handled by \texttt{FCLoop\-Find\-Topology\-Mappings}. The function takes a list of \texttt{FCTopology} objects as an input and searches for equivalent integral topologies among the elements of this list. In the case of success, it returns a set of mapping rules that contain loop momentum shifts and replacement rules for the \texttt{GLI}s. Using the option \texttt{PreferredTopologies} one can specify a set of target topologies that should be primarily considered during the construction of the mapping rules. Notice that in its current form \texttt{FCLoopFindTopologyMappings} can only find equivalent topologies with the same number of propagators. However, in real life calculations one often encounters topologies that do not have enough propagators to form a basis. Such incomplete topologies often fit into larger complete topologies, sometimes also denoted as \emph{supertopologies}. This issue can be addressed in the current development snapshot of \textsc{FeynCalc} as follows. Provided that one has some candidate supertopologies, one would first employ \texttt{FCLoopFindSubtopologies} to identify all nonvanishing subtopologies of the given supertopology. Then, one would add the list of subtopologies to the list of preferred topologies when running \texttt{FCLoopFindTopologyMappings}. Finally, using the function \texttt{FCLoopCreateRuleGLIToGLI} one can generate replacement rules that would eliminate the subtopologies in favor of the corresponding supertopology.

To apply this machinery to amplitudes \textsc{FeynCalc} also features two dedicated functions called \texttt{FCLoopFindTopologies} and \texttt{FCLoopApplyTopologyMappings}. The former identifies all distinct topologies present in the given amplitude, while the latter applies the mappings uncovered by \texttt{FCLoop\-FindTopology\-Mappings} to this amplitude. However, for practical reasons we would recommend against the idea of doing multiloop calculations entirely in \textsc{Mathematica}. The number of intermediate expressions in such computations can easily go into hundreds of thousands or even millions of terms even for a single diagram. Unlike \textsc{FORM}~\cite{Kuipers:2012rf}, \textsc{Mathematica} is simply not optimized to handle that level of complexity and any attempts to ignore this fact would most likely result into frustration as well as wasted time and efforts. The proper way to approach such calculations is to use \textsc{QGRAF}~\cite{Nogueira:1991ex} and \textsc{FORM}. Having extracted the list of distinct topologies from \textsc{FORM} one can readily employ \textsc{FeynCalc} to carry out the topology identification and then export the obtained mappings back into \textsc{FORM}. This is essentially how we envisage the usage of the new multiloop capabilities of \textsc{FeynCalc} in loop calculations. Of course, one might also think of situations where it should be admissible to  do the full calculation in a single notebook. One of such scenarios would be  asymptotic~\cite{Beneke:1997zp} expansions of single loop integrals in conjunction with \textsc{asy}~\cite{Jantzen:2012mw} and \textsc{FeynHelpers}~\cite{Shtabovenko:2016whf}. Here we would like to refer to the upcoming version of the \textsc{FeynHelpers}~\cite{Shtabovenko:FH} extension that would feature dedicated routines for automatizing such expansions.

\section{Master integrals}
\vspace{0.2cm}
After having completed all relevant IBP reductions, one usually starts to analyze the resulting master integrals. One of the first things to do is to create a list of master integrals from all integral families and check whether all of them are distinct. Just as in the case of equivalent topologies, equivalent integrals often cannot be recognized by eye. Here \textsc{FeynCalc} offers a function called \texttt{FCLoopFind\-Integral\-Mappings} that works similarly to \texttt{FindRules} in \textsc{FIRE}. The \texttt{PreferredIntegrals} option allows to map (if possible) the given list of integrals to a set of some predefined master integrals.

When presenting results from new multiloop calculations to the community, it is customary to visualize the relevant master integrals in form of graphs with styled edges. For example, edges representing massless propagators are often drawn as dashed lines, while a solid line stands for a massive denominator and a dot or a cross means that the corresponding propagator appears raised to an integer power (usually squared). Owing to the fact that the reconstruction of the graph representation from the propagator representation is not entirely trivial, it may take some time and effort to generate such figures for a new set of integrals. This task can be partially automatized using the functionality available in  \textsc{AZURITE}~\cite{Georgoudis:2016wff}, \textsc{PlanarityTest}~\cite{Bielas:2013rja}, \textsc{LiteRed}~\cite{Lee:2012cn,Lee:2013mka} and now also \textsc{FeynCalc}. Our implementation consists of two functions called \texttt{FCLoopIntegralToGraph} (for obtaining the graph) and \texttt{FCLoopGraphPlot} (for plotting the graph). Notice that the output of \texttt{FCLoopIntegralToGraph} can be, in principle, visualized using other tools \eg \textsc{GraphViz}~\cite{url:Graphviz}.

When it comes to the analytic evaluation of master integrals, currently the methods of differential equations~\cite{Kotikov:1991pm,Kotikov:1990kg,Kotikov:1991hm,Bern:1993kr,Remiddi:1997ny,Gehrmann:1999as} and Mellin-Barnes~\cite{Smirnov:1999gc,Tausk:1999vh,Anastasiou:2005cb,Czakon:2005rk} seem to be the most popular techniques to attack this problem. Yet in this report we would like to advocate another method that often turns out to be very successful when applied to integrals with only few scales. What is meant is the direct analytic integration starting from the Feynman parameter representation of the integral. In this context we would like to mention the \textsc{Maple} package \textsc{HyperInt}~\cite{Panzer:2014caa} that currently constitutes the most advanced publicly available tool for automatizing such integrations. Multiscale integrals often might require a suitable variable transformation to eliminate square roots that hinder the integration sequence. To this end we found the package \textsc{RationalizeRoots}~\cite{Besier:2019kco} (available both for \textsc{Maple} and \textsc{Mathematica}) to be very handy. Let us also remark that the results obtained by \textsc{HyperInt} in terms of complicated Goncharov Polylogarithms (GPLs)~\cite{Goncharov:1998kja} often can be further simplified  using the packages \textsc{HyperLogProcedures}~\cite{Schnetz:HLP} and \textsc{PolyLogTools}~\cite{Duhr:2019tlz}.

\textsc{FeynCalc} function \texttt{FCFeynmanParametrize} can obtain Feynman parametric representation for a wide range of loop integrals with quadratic or eikonal propagators. Possible types of supported integrals range from expressions with unit numerators over integrals with scalar products to tensor integrals with open indices. Furthermore,  \texttt{FCFeynmanParametrize} can handle not only Minkowskian but also Cartesian and Euclidean integrals, which essentially covers almost everything one might encounter in various Standard Model and EFT calculations. In addition to that, the function \texttt{FCFeynmanParameterJoin} allows the user to apply Feynman's formula for joining denominators sequentially. This means that instead of joining all propagators at once one may also split them into smaller subsets and join the propagators in those subsets first. The joining of the resulting propagators stemming from each subset is then done at the very last step. This procedure generates several sets of Feynman parameter variables that can be often integrated in a simpler way owing to the additional freedom when exploiting the Cheng-Wu~\cite{Cheng:1987ga} theorem. In our experience, choosing a smart way to join the propagators sequentially often can enable an analytic integration that would otherwise seem hopeless without introducing a proper sequence of Feynman variable transformations.

\section{Summary}
\vspace{0.2cm}
We presented a set of new \textsc{FeynCalc} functions that improve the usefulness of the package in multiloop calculations. The ideas and algorithms behind the routines are readily available in the  literature~\cite{Pak:2011xt,Hoff:2015kub,Jahn:2020tpj} and we gratefully acknowledge that the new way of dealing with multiloop integrals in the package draws many inspirations from such well-established tools as \textsc{FIRE}, \textsc{FIESTA}, \textsc{LiteRed}, \textsc{TopoID} and \textsc{pySecDec}. Nevertheless, we believe that our implementation of the relevant methods should be very useful not only to the existing \textsc{FeynCalc} users but also to the whole particle physics community. Although \textsc{FeynCalc} 10 has not yet been officially released, all the functions described in this report are publicly available through the so-called \emph{development version} of the package\footnote{\url{https://github.com/FeynCalc/feyncalc/wiki/Installation\#dev\_automatic\_installation}}. We would also like to draw the attention of the reader to the new documentation system\footnote{\url{https://feyncalc.github.io/referenceDev}} that already includes descriptions of all new functions and features helpful examples. 

In general, it is very gratifying to see the progress in the functionality of the package as compared to the situation some years ago when \textsc{FeynCalc} 9 was announced during the ACAT 2017~\cite{Shtabovenko:2016olh}. It would be far from correct, however, to claim that the package is now feature complete. Two main directions we would like to see \textsc{FeynCalc} go are helicity amplitude methods and a \textsc{FORM}-based library for computationally heavy tasks. Only time will tell to which extent those can be realized in future versions of the program.

\section*{Acknowledgements}
\vspace{0.2cm}
The research of V.\,S. was supported by the DFG under grant 396021762 -- TRR 257 ``Particle Physics Phenomenology after the Higgs Discovery.''. The author would like to thank Matthias Steinhauser and Marvin Gerlach for many useful discussions on different aspects of multiloop calculations and to Erik Panzer for his friendly and patient support on the usage of \textsc{HyperInt}.

\section*{References}
\bibliographystyle{iopart-num}
\bibliography{feyncalc.bib}

\providecommand{\newblock}{}
\begin{thebibliography}{10}
\expandafter\ifx\csname url\endcsname\relax
  \def\url#1{{\tt #1}}\fi
\expandafter\ifx\csname urlprefix\endcsname\relax\def\urlprefix{URL }\fi
\providecommand{\eprint}[2][]{\url{#2}}

\bibitem{Tancredi:2021oiq}
Tancredi L 2021 {\em {European Physical Society Conference on High Energy
  Physics 2021}\/} (\textit{Preprint} \eprint{2111.00205})

\bibitem{Mertig:1990an}
Mertig R, Bohm M and Denner A 1991 {\em Comput. Phys. Commun.\/} {\bf 64}
  345--359

\bibitem{Shtabovenko:2016sxi}
Shtabovenko V, Mertig R and Orellana F 2016 {\em Comput. Phys. Commun.\/} {\bf
  207} 432--444 (\textit{Preprint} \eprint{1601.01167})

\bibitem{Shtabovenko:2020gxv}
Shtabovenko V, Mertig R and Orellana F 2020 {\em Comput. Phys. Commun.\/} {\bf
  256} 107478 (\textit{Preprint} \eprint{2001.04407})

\bibitem{Brambilla:2020fla}
Brambilla N, Chung H~S, Shtabovenko V and Vairo A 2020 {\em JHEP\/} {\bf 11}
  130 (\textit{Preprint} \eprint{2006.15451})

\bibitem{Brambilla:2017kgw}
Brambilla N, Chen W, Jia Y, Shtabovenko V and Vairo A 2018 {\em Phys. Rev. D\/}
  {\bf 97} 096001 [Erratum: Phys.Rev.D 101, 039903 (2020)] (\textit{Preprint}
  \eprint{1712.06165})

\bibitem{Brambilla:2019fmu}
Brambilla N, Chung H~S, Lai W~K, Shtabovenko V and Vairo A 2019 {\em Phys. Rev.
  D\/} {\bf 100} 054038 (\textit{Preprint} \eprint{1907.06473})

\bibitem{Assi:2020srd}
Assi B and Kniehl B~A 2020  (\textit{Preprint} \eprint{2011.06437})

\bibitem{Assi:2020piz}
Assi B and Kniehl B~A 2020  (\textit{Preprint} \eprint{2011.06447})

\bibitem{Urban:2021evz}
Urban K 2021 {\em {Large Electroweak Corrections to Heavy WIMP Dark Matter
  Annihilation and Resummation}\/} Ph.D. thesis Munich, Tech. U.

\bibitem{Biondini:2021ccr}
Biondini S and Shtabovenko V 2021 {\em JHEP\/} {\bf 08} 114 (\textit{Preprint}
  \eprint{2106.06472})

\bibitem{Chung:2021efj}
Chung H~S 2021 {\em JHEP\/} {\bf 09} 195 (\textit{Preprint}
  \eprint{2106.15514})

\bibitem{Frugiuele:2021bic}
Frugiuele C and Peset C 2021  (\textit{Preprint} \eprint{2107.13512})

\bibitem{Biondini:2021ycj}
Biondini S and Shtabovenko V 2021  (\textit{Preprint} \eprint{2112.10145})

\bibitem{Shtabovenko:FCX}
Shtabovenko V, Mertig R and Orellana F {\em in preparation\/}

\bibitem{Chetyrkin:1981qh}
Chetyrkin K~G and Tkachov F~V 1981 {\em Nucl. Phys. B\/} {\bf 192} 159--204

\bibitem{Tkachov:1981wb}
Tkachov F~V 1981 {\em Phys. Lett. B\/} {\bf 100} 65--68

\bibitem{Dixon:2018qgp}
Dixon L~J, Luo M~X, Shtabovenko V, Yang T~Z and Zhu H~X 2018 {\em Phys. Rev.
  Lett.\/} {\bf 120} 102001 (\textit{Preprint} \eprint{1801.03219})

\bibitem{Luo:2019nig}
Luo M~X, Shtabovenko V, Yang T~Z and Zhu H~X 2019 {\em JHEP\/} {\bf 06} 037
  (\textit{Preprint} \eprint{1903.07277})

\bibitem{Gao:2020vyx}
Gao J, Shtabovenko V and Yang T~Z 2021 {\em JHEP\/} {\bf 02} 210
  (\textit{Preprint} \eprint{2012.14188})

\bibitem{Gerlach:2021xtb}
Gerlach M, Nierste U, Shtabovenko V and Steinhauser M 2021 {\em JHEP\/} {\bf
  07} 043 (\textit{Preprint} \eprint{2106.05979})

\bibitem{Studerus:2009ye}
Studerus C 2010 {\em Comput. Phys. Commun.\/} {\bf 181} 1293--1300
  (\textit{Preprint} \eprint{0912.2546})

\bibitem{vonManteuffel:2012np}
von Manteuffel A and Studerus C 2012  (\textit{Preprint} \eprint{1201.4330})

\bibitem{Smirnov:2019qkx}
Smirnov A~V and Chuharev F~S 2020 {\em Comput. Phys. Commun.\/} {\bf 247Â }
  106877 (\textit{Preprint} \eprint{1901.07808})

\bibitem{Maierhofer:2017gsa}
Maierh\"ofer P, Usovitsch J and Uwer P 2018 {\em Comput. Phys. Commun.\/} {\bf
  230} 99--112 (\textit{Preprint} \eprint{1705.05610})

\bibitem{Klappert:2020nbg}
Klappert J, Lange F, Maierh\"ofer P and Usovitsch J 2021 {\em Comput. Phys.
  Commun.\/} {\bf 266} 108024 (\textit{Preprint} \eprint{2008.06494})

\bibitem{Lee:2012cn}
Lee R~N 2012  (\textit{Preprint} \eprint{1212.2685})

\bibitem{Lee:2013mka}
Lee R~N 2014 {\em J. Phys. Conf. Ser.\/} {\bf 523} 012059 (\textit{Preprint}
  \eprint{1310.1145})

\bibitem{Harlander:1998cmq}
Harlander R, Seidensticker T and Steinhauser M 1998 {\em Phys. Lett. B\/} {\bf
  426} 125--132 (\textit{Preprint} \eprint{hep-ph/9712228})

\bibitem{Seidensticker:1999bb}
Seidensticker T 1999 {\em {6th International Workshop on New Computing
  Techniques in Physics Research: Software Engineering, Artificial Intelligence
  Neural Nets, Genetic Algorithms, Symbolic Algebra, Automatic Calculation}\/}
  (\textit{Preprint} \eprint{hep-ph/9905298})

\bibitem{Hoff:2016pot}
Hoff J 2016 {\em J. Phys. Conf. Ser.\/} {\bf 762} 012061 (\textit{Preprint}
  \eprint{1607.04465})

\bibitem{url:Feynson}
 {\em \url{https://github.com/magv/feynson/}\/}

\bibitem{Gerlach:tapir}
Gerlach M, Herren F and Lang M {\em in preparation\/}

\bibitem{Borowka:2017idc}
Borowka S, Heinrich G, Jahn S, Jones S~P, Kerner M, Schlenk J and Zirke T 2018
  {\em Comput. Phys. Commun.\/} {\bf 222} 313--326 (\textit{Preprint}
  \eprint{1703.09692})

\bibitem{Borowka:2018goh}
Borowka S, Heinrich G, Jahn S, Jones S~P, Kerner M and Schlenk J 2019 {\em
  Comput. Phys. Commun.\/} {\bf 240} 120--137 (\textit{Preprint}
  \eprint{1811.11720})

\bibitem{Heinrich:2021dbf}
Heinrich G, Jahn S, Jones S~P, Kerner M, Langer F, Magerya V, P\"oldaru A,
  Schlenk J and Villa E 2021  (\textit{Preprint} \eprint{2108.10807})

\bibitem{Pak:2011xt}
Pak A 2012 {\em J. Phys. Conf. Ser.\/} {\bf 368} 012049 (\textit{Preprint}
  \eprint{1111.0868})

\bibitem{Hoff:2015kub}
Hoff J~S 2015 {\em {Methods for multiloop calculations and Higgs boson
  production at the LHC}\/} Ph.D. thesis KIT, Karlsruhe

\bibitem{Bogner:2010kv}
Bogner C and Weinzierl S 2010 {\em Int. J. Mod. Phys. A\/} {\bf 25} 2585--2618
  (\textit{Preprint} \eprint{1002.3458})

\bibitem{Smirnov:2012gma}
Smirnov V~A 2012 {\em {Analytic tools for Feynman integrals}\/} vol 250

\bibitem{Jahn:2020tpj}
Jahn S 2020 {\em {Automation of Multi-Loop Amplitude Calculations}\/} Ph.D.
  thesis Munich, Tech. U.

\bibitem{Smirnov:2021rhf}
Smirnov A~V, Shapurov N~D and Vysotsky L~I 2021  (\textit{Preprint}
  \eprint{2110.11660})

\bibitem{Kuipers:2012rf}
Kuipers J, Ueda T, Vermaseren J~A~M and Vollinga J 2013 {\em Comput. Phys.
  Commun.\/} {\bf 184} 1453--1467 (\textit{Preprint} \eprint{1203.6543})

\bibitem{Nogueira:1991ex}
Nogueira P 1993 {\em J. Comput. Phys.\/} {\bf 105} 279--289

\bibitem{Beneke:1997zp}
Beneke M and Smirnov V~A 1998 {\em Nucl. Phys. B\/} {\bf 522} 321--344
  (\textit{Preprint} \eprint{hep-ph/9711391})

\bibitem{Jantzen:2012mw}
Jantzen B, Smirnov A~V and Smirnov V~A 2012 {\em Eur. Phys. J. C\/} {\bf 72}
  2139 (\textit{Preprint} \eprint{1206.0546})

\bibitem{Shtabovenko:2016whf}
Shtabovenko V 2017 {\em Comput. Phys. Commun.\/} {\bf 218} 48--65
  (\textit{Preprint} \eprint{1611.06793})

\bibitem{Shtabovenko:FH}
Shtabovenko V {\em in preparation\/}

\bibitem{Georgoudis:2016wff}
Georgoudis A, Larsen K~J and Zhang Y 2017 {\em Comput. Phys. Commun.\/} {\bf
  221} 203--215 (\textit{Preprint} \eprint{1612.04252})

\bibitem{Bielas:2013rja}
Bielas K, Dubovyk I, Gluza J and Riemann T 2013 {\em Acta Phys. Polon. B\/}
  {\bf 44} 2249--2255 (\textit{Preprint} \eprint{1312.5603})

\bibitem{url:Graphviz}
 {\em \url{https://www.graphviz.org/}\/}

\bibitem{Kotikov:1991pm}
Kotikov A~V 1991 {\em Phys. Lett. B\/} {\bf 267} 123--127 [Erratum: Phys.Lett.B
  295, 409--409 (1992)]

\bibitem{Kotikov:1990kg}
Kotikov A~V 1991 {\em Phys. Lett. B\/} {\bf 254} 158--164

\bibitem{Kotikov:1991hm}
Kotikov A~V 1991 {\em Phys. Lett. B\/} {\bf 259} 314--322

\bibitem{Bern:1993kr}
Bern Z, Dixon L~J and Kosower D~A 1994 {\em Nucl. Phys. B\/} {\bf 412} 751--816
  (\textit{Preprint} \eprint{hep-ph/9306240})

\bibitem{Remiddi:1997ny}
Remiddi E 1997 {\em Nuovo Cim. A\/} {\bf 110} 1435--1452 (\textit{Preprint}
  \eprint{hep-th/9711188})

\bibitem{Gehrmann:1999as}
Gehrmann T and Remiddi E 2000 {\em Nucl. Phys. B\/} {\bf 580} 485--518
  (\textit{Preprint} \eprint{hep-ph/9912329})

\bibitem{Smirnov:1999gc}
Smirnov V~A 1999 {\em Phys. Lett. B\/} {\bf 460} 397--404 (\textit{Preprint}
  \eprint{hep-ph/9905323})

\bibitem{Tausk:1999vh}
Tausk J~B 1999 {\em Phys. Lett. B\/} {\bf 469} 225--234 (\textit{Preprint}
  \eprint{hep-ph/9909506})

\bibitem{Anastasiou:2005cb}
Anastasiou C and Daleo A 2006 {\em JHEP\/} {\bf 10} 031 (\textit{Preprint}
  \eprint{hep-ph/0511176})

\bibitem{Czakon:2005rk}
Czakon M 2006 {\em Comput. Phys. Commun.\/} {\bf 175} 559--571
  (\textit{Preprint} \eprint{hep-ph/0511200})

\bibitem{Panzer:2014caa}
Panzer E 2015 {\em Comput. Phys. Commun.\/} {\bf 188} 148--166
  (\textit{Preprint} \eprint{1403.3385})

\bibitem{Besier:2019kco}
Besier M, Wasser P and Weinzierl S 2020 {\em Comput. Phys. Commun.\/} {\bf 253}
  107197 (\textit{Preprint} \eprint{1910.13251})

\bibitem{Goncharov:1998kja}
Goncharov A~B 1998 {\em Math. Res. Lett.\/} {\bf 5} 497--516 (\textit{Preprint}
  \eprint{1105.2076})

\bibitem{Schnetz:HLP}
Schnetz O {\em \url{https://www.math.fau.de/person/oliver-schnetz}\/}

\bibitem{Duhr:2019tlz}
Duhr C and Dulat F 2019 {\em JHEP\/} {\bf 08} 135 (\textit{Preprint}
  \eprint{1904.07279})

\bibitem{Cheng:1987ga}
Cheng H and Wu T~T 1987 {\em {EXPANDING PROTONS: SCATTERING AT
  HIGH-ENERGIES}\/}

\bibitem{Shtabovenko:2016olh}
Shtabovenko V 2016 {\em J. Phys. Conf. Ser.\/} {\bf 762} 012064
  (\textit{Preprint} \eprint{1604.06709})

\end{thebibliography}

\end{document}